\documentclass[12pt,pdf,epsfig,color,png]{article}
\usepackage{amsmath}
\usepackage{amsfonts}
\usepackage{amsmath}
\usepackage{slashed}
\usepackage{amsxtra}
\usepackage{amstext}
\usepackage{amssymb}
\usepackage{color}
\usepackage{float}
\usepackage{graphicx}
\usepackage{subcaption}
\numberwithin{equation}{section}
\usepackage[dvips]{epsfig}
\usepackage{appendix}
\usepackage{youngtab}
\allowdisplaybreaks

\newcommand{\be}{\begin{equation}}
\newcommand{\ee}{\end{equation}}
\newcommand{\beq}{\begin{equation}}
\newcommand{\eeq}{\end{equation}}
\newcommand{\ba}{\begin{eqnarray}}
\newcommand{\ea}{\end{eqnarray}}
\newcommand{\bea}{\begin{eqnarray}}
\newcommand{\eea}{\end{eqnarray}}

\begin{document}
\baselineskip=15.5pt \pagestyle{plain} \setcounter{page}{1}
%
\begin{titlepage}

\vskip 0.8cm

\begin{center}
%
%
%
%
%

{\Large \bf Glueball-fermion hard scattering from type IIB superstring theory}

\vskip 1.cm

{\large {{\bf Lucas Martin{\footnote{\tt lucasmartinar@iflp.unlp.edu.ar}}, Martin Parlanti{\footnote{\tt martin.parlanti@fisica.unlp.edu.ar}}} {\bf and  Martin
Schvellinger}{\footnote{\tt martin@fisica.unlp.edu.ar}}}}

\vskip 1.cm

{\it Instituto de F\'{\i}sica La Plata-UNLP-CONICET. \\
Boulevard 113 e 63 y 64, (1900) La Plata, Buenos Aires, Argentina \\
and \\
Departamento de F\'{\i}sica, Facultad de Ciencias Exactas,
Universidad Nacional de La Plata. \\
Calle 49 y 115, C.C. 67, (1900) La Plata, Buenos Aires, Argentina.}

\vspace{1.cm}

{\bf Abstract}

\end{center}

\vspace{.5cm}

We calculate the glueball-fermion hard-scattering amplitude from the dilaton-dilatino closed string scattering amplitude in type IIB superstring theory, in the framework of the gauge/string theory duality. We investigate its high-energy scaling at fixed angle and also in the Regge limit. We derive the leading and sub-leading terms contributing to the scattering cross section. This dual calculation of the glueball-fermion scattering amplitude is particularly interesting since it involves the scattering of two different types of external states. We calculate explicitly some angular integrals for two scalar spherical harmonics and two spinor spherical harmonics on the five-sphere, leading to selection rules.

\noindent

\end{titlepage}

\newpage

{\small \tableofcontents}

\newpage

%
%
\section{Introduction}
%
%

In reference \cite{Martin:2024jpe} we initiated a systematic study of elastic hard scattering processes at fixed angle and also considering the Regge limit, in terms of type IIB superstring theory scattering amplitudes. Since it is based on the gauge/string theory duality, this approach in principle holds for confining non-Abelian gauge theories at strong coupling and in the limit of large number of color degrees of freedom. The idea was to extend the calculation proposed by Polchinski and Strassler for $2 \rightarrow m$ glueballs hard scattering, using type IIB superstring theory scattering amplitudes at tree level \cite{Polchinski:2001tt}, to the case of four spin-1/2 fermions \cite{Martin:2024jpe}. This extension is not straightforward since it requires the full explicit form of the four-dilatino closed-string scattering amplitude, which we obtained in \cite{Martin:2025pug} through a complicated algebraic calculation by using the Kawai-Lewellen-Tye (KLT) relations \cite{Kawai:1985xq}. After integrating in the radial coordinate as well as the five angular coordinates we obtained an effective four-dimensional scattering amplitude which in the high-energy limit, leads to two important results \cite{Martin:2024jpe}. Firstly, we found that the connected invariant scattering amplitude of four spin-1/2 fermions at fixed scattering angle has a dominant contribution proportional to $s^{2-\tau/2}$, where $s$ is the four-dimensional $s$-channel Mandelstam variable and $\tau=\sum_{j=1}^{4}\tau_j$. Notice that $\tau_j$ denotes the twist of the operator  which creates the $j$-th fermionic state of the dual gauge theory. The second very interesting result of \cite{Martin:2024jpe} is the Regge behavior which we obtained for $s \gg |t|$ (where $t$ is the four-dimensional $t$-channel Mandelstam variable). Thus, it is possible to explicitly calculate the high-energy scaling of the gauge theory scattering amplitude involving four spin-1/2 fermions from the four-dilatino type IIB superstring theory scattering amplitude.

Scaling laws for large-momentum-transfer scattering processes have been investigated since long time ago. In particular, fixed-angle exclusive and inclusive scattering amplitudes, which can be derived from renormalizable field theories with asymptotically scale-invariant interactions and hadronic wave functions finite at the origin, were studied in detail in references \cite{Brodsky:1973kr,Brodsky:1974vy} for non-Abelian gauge field theories. Our results of reference \cite{Martin:2024jpe}, which involve the elastic scattering of four spin-1/2 fermions, are consistent with the old results reported in \cite{Brodsky:1973kr,Brodsky:1974vy}. In the present work we investigate the high-energy scattering process involving two different particle species, namely: the glueball-fermion elastic scattering. We would like to emphasize that, to our knowledge, this is the first calculation of the glueball-fermion elastic scattering reported in the literature, at least in the context of the gauge/string theory duality. It may be useful for further studies.

The starting point is the dilaton-dilatino scattering amplitude in ten-dimensional Min\-kows\-ki spacetime which we have recenclty derived within type II superstring theory for closed strings \cite{Martin:2025ije} using the KLT relations. Then, we integrate in the radial coordinate and in the five-sphere coordinates, obtaining the glueball-fermion scattering amplitude in four dimensions. The quantum numbers of the glueball and the spin-1/2 fermion scattered states remain unaltered after interaction. We show how this selection rule results in the case of the scattering of ${\cal {N}}=4$ supersymmetric Yang-Mills theory (SYM) states created by single color-trace\footnote{This refers to single-trace operators for which the trace is taken on the adjoint representation of the $SU(N_c)$ color gauge group.} operators which are dual to the Kaluza-Klein states emerging from the spontaneous compactification of type IIB supergravity on AdS$_5 \times S^5$ \cite{Kim:1985ez} (see also \cite{vanNieuwenhuizen:2012zk,vanNieuwenhuizen:2019lbe}). By explicit calculation we show that our present results are consistent with those of reference \cite{Brodsky:1974vy} for two different species of particles at fixed angle. In particular, we obtain that the differential cross section for the hard scattering of two glueballs and two spin-1/2 fermions with minimal twists, i.e. $\tau_{\text{glueball}}=4$ and $\tau_{\text{spin-1/2 fermion}}=3$, is $d\sigma/dt \propto s^{-11} f(\cos(\theta))$, where we have included a function of the scattering angle $\theta$. We also have obtained the Regge behavior of this scattering amplitude.

Another very interesting aspect of this calculation is that since the expression of the ten-dimensional dilaton-dilatino scattering amplitude that we derived in reference \cite{Martin:2025ije} is very compact, the calculations which we have developed in the present work give the full result for the glueball-fermion hard scattering amplitude in four dimensions. In our previous work \cite{Martin:2024jpe} in the case of four spin-1/2 fermions scattering we only were able to carry out the calculations using some representative terms of the ten-dimensional four-dilatino scattering amplitude, since the expression that we derived for it in \cite{Martin:2025pug} is by far much more complicated in comparison with the case of two dilatons and two dilatinos which we have recently obtained in \cite{Martin:2025ije}.

In previous papers some aspects of the partonic behavior associted with the holographic dual calculation of strongly coupled gauge theories have been investigated. For instance, hard scattering processes of glueballs from different holographic dual models \cite{Karch:2006pv}, and holographic mesons \cite{Witten:1998qj,Sakai:2004cn} have been studied in \cite{Bianchi:2021sug}. Also, proton-proton and proton-anti proton scattering processes have been investigated in \cite{Domokos:2009hm,Domokos:2010ma}. More recently, in reference \cite{Armoni:2024gqv} the high-energy fixed-angle scattering of pions and $\rho$ mesons has been studied in a bottom-up holographic dual QCD model. Since from the gauge field theory perspective it is possible to consider scaling behavior of inclusive processes like deep inelastic scattering (DIS) \cite{Brodsky:1973kr,Brodsky:1974vy}, it is also worth mentioning some developments carried out following reference \cite{Polchinski:2002jw}. For instance, DIS of charged leptons off spin-1/2 fermions has been investigated in \cite{Kovensky:2018xxa} in the context of type IIB supergravity on AdS$_5 \times S^5$. In a series of papers 
\cite{Koile:2011aa,Koile:2013hba,Jorrin:2016rbx,Jorrin:2016ccr}, using the top-down holographic dual description of DIS processes of charged leptons from holographic pseudo-scalar mesons and vector mesons have been investigated. Moreover, in reference \cite{Brower:2006ea} it has been proposed the so-called BPST Pomeron obtained from the reggeization of the graviton in AdS$_5 \times S^5$, which allows to calculate the proton structure function $F_2$ \cite{Brower:2010wf}, leading to an excellent level of agreement in comparison with experimental data \cite{Brower:2010wf,Jorrin:2022lua}.
In addition, in \cite{Kovensky:2018xxa} it has been developed the holographic $A_4$ Pomeron, which exchanges a single Reggeized vector field. The holographic $A_4$ Pomeron fits experimental data very well for the proton antisymmetric structure function $g_1$ \cite{Jorrin:2022lua,Borsa:2023tqr} using a single free parameter, leading to important predictions for the forthcoming Electron-Ion Collider \cite{AbdulKhalek:2021gbh}.

This work is organized as follows. In section 2 we briefly discuss the full dilaton-dilatino scattering amplitude which we have obtained in \cite{Martin:2025ije} from type II superstring theory. This is the starting point of the calculation. In section 3 we present the calculation of the glueball-fermion scattering amplitude from the dilaton-dilatino scattering amplitude. We introduce the basic setup, where we discuss the philosophy of the approach based on the Polchinski-Strassler original proposal for glueball hard scattering. Then, we introduce the dilaton and dilatino wave functions, and obtain the scattering amplitude for a spin-1/2 fermion and a glueball in four dimensions. In section 4 we investigated the Regge limit of the glueball-fermion scattering amplitude. In section 5 we discuss our results, describing the  ${\cal {N}}=4$ SYM operators which create the glueballs and spin-1/2 fermionic states, and their relation to the Kaluza-Klein states from type IIB supergravity on AdS$_5 \times S^5$. We also comment on the scalar and spinor spherical harmonics on $S^5$, as well as selection rules for this scattering process. We compare our results for the scattering cross section with those of Brodsky and Farrar obtained in the framework of confining quantum field theory.

%
%
\section{Dilaton-dilatino scattering amplitude from type IIB superstring theory}
%
%

The main idea of the present work is to calculate the glueball-fermion hard-scattering amplitude and study its scaling behavior at fixed angle and in the Regge limit. The starting point is the dilaton-dilatino closed-string scattering amplitude of the form $\text{NS-NS}+\text{NS-R}\rightarrow \text{NS-NS}+\text{NS-R}$ in type II superstring theory calculated in \cite{Martin:2025ije}.

Let us very briefly recall how the KLT relations work in the present case. The four-point closed-string scattering amplitude in ten-dimensional Minkowski spacetime can be written as: 
\begin{eqnarray}
\mathcal{A}_{\text{string}}^{10d}(\tilde{p}) = 4 g_s^2 \alpha'^3 F(\tilde{p}\sqrt{\alpha^\prime}) \, , \label{first-amplitude-10d}
\end{eqnarray}
where the function $F(\tilde{p}\sqrt{\alpha^\prime})$ contains the pole structure of the scattering amplitude as well as the kinematic factor:
\begin{eqnarray}
F(\tilde{p} \sqrt{\alpha'}) = \prod_{\chi=\tilde{s},\tilde{t},\tilde{u}} \frac{\Gamma(-\alpha' \chi/4)}{\Gamma(1+\alpha' \chi/4)} \ K(\tilde{p} \sqrt{\alpha'}) \, . \label{FGammaK} 
\end{eqnarray}
The ten-dimensional Mandelstam variables are:
\begin{equation}
\tilde{s}=-(k_1+k_2)^2 \, , \,\,\,\,\,\,\, 
\tilde{t}=-(k_1+k_4)^2 \, , \,\,\,\,\,\,\,
\tilde{u}=-(k_1+k_3)^2 \, , 
\end{equation}
where the four ten-dimensional incoming momenta are indicated as $k_i$'s. The function $K(\tilde{p}\sqrt{\alpha'})$ is defined in terms of the closed-string kinematic factor  $K_{\text{closed}}^{\text{dilaton-dilatino}}(\tilde{k}_1,k_2,k_3,\tilde{k}_4)$ as follows:
\begin{equation}
K(\tilde{p}\sqrt{\alpha'})= - \pi \alpha'^4 K_{\text{closed}}^{\text{dilaton-dilatino}}(\tilde{k}_1,k_2,k_3,\tilde{k}_4) \, . \label{K2-9}
\end{equation}
Ten-dimensional momenta of external fermionic states in the scattering amplitudes are indicated with tildes. By using directly the KLT relations the dilaton-dilatino kinematic factor can be expressed as the tensor product of two kinematic factors corresponding to four-open string scattering amplitudes:
\begin{eqnarray}
&& K_{\text{closed}}^{\text{dilaton-dilatino}}(\tilde{k}_1, k_2, k_3, \tilde{k}_4)  \nonumber \\
&&  \,\,\,\,\,\,\,\,\, =  K_{\text{open}}^{\text{bosonic}}(k_1/2, k_2/2, k_3/2, k_4/2) \otimes  K_{\text{open}}^{\text{fermionic}}(\tilde{k}_1/2, k_2/2, k_3/2,\tilde{k}_4/2) \,  ,\label{KLT-1} 
\end{eqnarray}
where we have considered that particles 3 and 4 can be exchanged as follows:
\begin{eqnarray}
K_{\text{open}}^{\text{fermionic}}(\tilde{k}_1/2, k_2/2,\tilde{k}_4/2, k_3/2) =
K_{\text{open}}^{\text{fermionic}}(\tilde{k}_1/2, k_2/2, k_3/2,\tilde{k}_4/2) \, .
\end{eqnarray}
The kinematic factor for the open-string scattering amplitude corresponding to four bosonic states is given by \cite{Green:1987sp,Green:1981xx}:
\begin{eqnarray}
K_{\text{open}}^{\text{bosonic}}(k_1/2, k_2/2, k_3/2, k_4/2) &=& -\frac{1}{4\cdot 16}(\tilde{s}\tilde{u}\zeta_1\cdot \zeta_4 \zeta_3\cdot \zeta_2+\tilde{s}\tilde{t}\zeta_1\cdot \zeta_3 \zeta_4\cdot \zeta_2 +\tilde{t}\tilde{u}\zeta_1\cdot \zeta_2 \zeta_4\cdot \zeta_3)\nonumber\\
    &+&\frac{1}{2 \cdot 16}\tilde{s}(\zeta_1\cdot k_4 \zeta_3\cdot k_2\zeta_4\cdot \zeta_2+\zeta_2\cdot k_3 \zeta_4\cdot k_1\zeta_1\cdot \zeta_3 \nonumber \\
    &+&\zeta_1\cdot k_3 \zeta_4\cdot k_2\zeta_3\cdot \zeta_2+\zeta_2\cdot k_4 \zeta_3\cdot k_1\zeta_1\cdot \zeta_4) \nonumber \\
    &+& \frac{1}{2\cdot 16}\tilde{t}(\zeta_2\cdot k_1 \zeta_4\cdot k_3\zeta_3\cdot \zeta_1+\zeta_3\cdot k_4 \zeta_1\cdot k_2\zeta_2\cdot \zeta_4 \nonumber \\
    &+&\zeta_2\cdot k_4 \zeta_1\cdot k_3\zeta_3\cdot \zeta_4+\zeta_3\cdot k_1 \zeta_4\cdot k_2\zeta_2\cdot \zeta_1)\nonumber\\
    &+& \frac{1}{2\cdot 16}\tilde{u}(\zeta_1\cdot k_2 \zeta_4\cdot k_3\zeta_3\cdot \zeta_2+\zeta_3\cdot k_4 \zeta_2\cdot k_1\zeta_1\cdot \zeta_4  \nonumber \\
    &+&\zeta_1\cdot k_4 \zeta_2\cdot k_3\zeta_3\cdot \zeta_4+\zeta_3\cdot k_2 \zeta_4\cdot k_1\zeta_1\cdot \zeta_2)  
 \, , \label{Kopen4}
\end{eqnarray}
where $\zeta_i$'s denote vector polarizations of the external open-string bosonic states.\footnote{Note that in reference \cite{Martin:2024jpe} there should be an overall factor 1/16 multiplying the right-hand side of its equation (2.11). It only modifies an overall numerical factor which does not alter the power behavior of the fixed-angle calculation nor the Regge limit of the four-glueball scattering amplitude. In the case of four-fermion scattering, obviously there are no changes.} In this factor we have reordered the Mandelstam variables in comparison with equation (7.4.42) of reference \cite{Green:1987sp}. With the purpose of making the notation simpler we define:
\begin{equation}
K_{\text{open}}^{\text{bosonic}}(1, 2, 3, 4)\equiv K_{\text{open}}^{\text{bosonic}}(k_1/2, k_2/2, k_3/2, k_4/2) \, , 
\end{equation}
and similarly:
\begin{equation}
K_{\text{open}}^{\text{fermionic}}(\tilde{1}, 2, 3,\tilde{4}) \equiv K_{\text{open}}^{\text{fermionic}}(\tilde{k}_1/2, k_2/2, k_3/2,\tilde{k}_4/2) \, ,
\end{equation}
where the open-string kinematic factor involving two bosons and two fermions is:
\begin{eqnarray}
K_{\text{open}}^{\text{fermionic}}(\tilde{1},2,3,\tilde{4})&=&-\frac{\tilde{t}}{16}\bar{u}_1\Gamma \cdot \zeta_2 \ \Gamma \cdot(k_3+k_4) \ \Gamma\cdot \zeta_3 u_4 + \frac{\tilde{s}}{8} [ \bar{u}_1\Gamma\cdot\zeta_3 u_4 \ k_3 \cdot \zeta_2  \nonumber \\
&& -\bar{u}_1\Gamma\cdot\zeta_2 u_4 \ k_2 \cdot \zeta_3-\bar{u}_1\Gamma\cdot k_3 u_4 \ \zeta_2 \cdot \zeta_3 ] \, . \label{kinematic-factor-two-fermions}
\end{eqnarray}
We have divided each ten-dimensional momentum by two, as required by the KLT relations. In addition, $\Gamma^M$ represents a $32 \times 32$-component Dirac gamma matrix in ten-dimensional Minkowski spacetime.

Next, we carry out the explicit tensor product (\ref{KLT-1}). It is important to consider that the product of two vector polarizations from the open-string kinematic factors leads to a second rank tensor defined as:
\begin{eqnarray}
    \Theta_i^{MM'}=\zeta_i^M\otimes \zeta_i^{M'} \, ,
\end{eqnarray}
where $M$ and $M'$ are ten-dimensional Lorentz indices. For dilatons we set $i=2, 3$. In particular, this tensor can be written in terms of the dilaton wave function $\Phi_i$ as follows:
\begin{eqnarray}
   \Theta_i^{MM'}=\frac{1}{ \sqrt{8}} \left(\eta^{MM'}-k_i^M\bar{k}_i^{M'}-k_i^{M'}\bar{k}_i^{M}\right)  \Phi_i \, ,
\end{eqnarray} 
being $\eta^{MM'}$ the ten-dimensional Minkowski spacetime metric. For each $i$-th external state there is an auxiliary ten-dimensional momentum $\bar{k}_i$, satisfying the following conditions: $k \cdot \bar{k}=1$ and $\bar{k} \cdot \bar{k}=0$ \cite{Gross:1986mw}.

The 32-component dilatino of type II superstring theory, $\lambda_i$, satisfies the following relation \cite{Garousi:1996ad}:
\begin{eqnarray}
    (\Gamma^M)^\alpha_\beta \lambda_i^\beta = u_i^\alpha \otimes \zeta_i^M   \, ,
\end{eqnarray}
where $u_i^\alpha$ is the $\alpha$ component of the spinor wave function of equation (\ref{kinematic-factor-two-fermions}). Notice that $\alpha$ and $\beta$ label spinor indices, and for dilatinos we set $i=1, 4$.

From the explicit tensor product in $K_{\text{closed}}^{\text{dilaton-dilatino}}(\tilde{1}, 2, 3, \tilde{4})$ defined in equation (\ref{KLT-1}) there are 60 terms. Fortunately, after some tedious algebra these terms can be summed, obtaining a very compact expression given by:
\begin{eqnarray}
&& K_{\text{closed}}^{\text{dilaton-dilatino}}(\tilde{1},2,3,\tilde{4}) =  \nonumber \\
&& \,\,\,\,\,\,\,\, \,\,\,\,\,\,\,\,\,\,\,\,\,\,\,\, \,\,\,\,\,\,\,\, \frac{ \left(8 \tilde{s}^3+\tilde{u} \left(16 \tilde{s}^2+86 \tilde{s} \tilde{t}+35 \tilde{t}^2\right)+50 \tilde{s}^2 \tilde{t}+5 \tilde{s} \tilde{t}^2-4 \tilde{u}^2 (4 \tilde{s}+\tilde{t})-10 \tilde{t}^3\right)}{2048} \nonumber \\
&& \,\,\,\,\,\,\,\, \,\,\,\,\,\,\,\,\,\,\,\,\,\,\,\, \,\,\,\,\,\,\,\, \times \ \bar{\lambda}_1\Phi_2\Phi_3(k_3\cdot \Gamma)\lambda_4 \, . \label{K4-dilaton-dilatino-full}
\end{eqnarray}
Using the expressions of the Mandelstam variables in terms of the scattering angle $\theta$:
\begin{equation}
\tilde{t}= -\frac{\tilde{s}}{2} (1-\cos\theta) \, , \,\,\,\,\,\,\,\, \text{and} \,\,\,\,\,\,\,\, \tilde{u}=-\frac{\tilde{s}}{2} (1+\cos\theta) \, , \label{kinematics} 
\end{equation}
equation (\ref{K4-dilaton-dilatino-full}) becomes:
\begin{eqnarray}
K_{\text{closed}}^{\text{dilaton-dilatino}}(\tilde{1},2,3,\tilde{4}) &=&  -\frac{\tilde{s}^3 (49 \cos (3 \theta)+266 \cos (2 \theta)-97 \cos (\theta)+550)}{65536} \nonumber \\
&& \times \ \bar{\lambda}_1\Phi_2\Phi_3(k_3\cdot \Gamma)\lambda_4 \, . \label{K4-dilaton-dilatino-full-theta} 
\end{eqnarray}
For the complete calculation of this ten-dimensional closed string scattering amplitude we refer the reader to our work introduced in reference \cite{Martin:2025ije}.

%
%
\section{Glueball-fermion scattering amplitude from dilaton-dilatino scattering amplitude at fixed angle}
%
%

In this section we calculate the four-dimensional glueball-fermion scattering amplitude in the gauge theory in four-dimensional Minkowski spacetime from the dilaton-dilatino scattering amplitude of type II superstring theory which we have discussed in the previous section. The idea is to extend the prescription developed by Polchinski and Strassler \cite{Polchinski:2001tt} for the high-energy limit of $2 \rightarrow m$ glueballs in confining gauge theories at fixed angle to the case of glueball-fermion to glueball-fermion hard scattering. Firstly, we introduce the basic setup. Secondly, we study the dilaton and dilatino wave functions. Then, we proceed to derive the high-energy behavior of the glueball-fermion scattering amplitude at fixed scattering angle.

\subsection{Basic setup}

The relation between these two physically very different scattering amplitudes is given by the following equation:
\begin{equation}
\mathcal{A}_{4}^{\text{glueball-fermion}}(p) = \int_{r_0}^{\infty} dr \int_{S^5} d\Omega_5 \ \sqrt{-g} \ \tilde{\mathcal{A}}_{\text{closed}}^{\text{dilaton-dilatino}}(\tilde{p})  \, . \label{A4}
\end{equation}
where $\mathcal{A}_{4}^{\text{glueball-fermion}}(p)$ is the SYM theory four-dimensional scattering amplitude corresponding to two glueballs and two spin-1/2 fermions scattering. In addition, $\mathcal{A}_{\text{closed}}^{\text{dilaton-dilatino}}(\tilde{p})$ 
is the closed-string scattering amplitude of type IIB superstring theory corresponding to the scattering of two dilatons and two dilatinos in ten-dimensional Minkowski spacetime. Thus, equation (\ref{A4}) is a natural extension to the case of the glueball-fermion scattering amplitude calculation from the Polchinski-Strassler proposal corresponding to four-glueball hard scattering \cite{Polchinski:2001tt}, and also from our previous calculation of four spin-1/2 fermions from type IIB string theory scattering amplitude of four dilatinos \cite{Martin:2024jpe}. The assumption made to arrive to this equation is that the string scattering process is localized in the bulk. Then, a coherent integration over all possible localizations of the scattering process at $(r, \Omega_5)$ is carried out. We use the definitions and construction presented in section 2 to write explicitly the four-dimensional scattering amplitude $\mathcal{A}_{4}^{\text{glueball-fermion}}(p)$. The fact that the background is AdS$_5 \times S^5$ implies that we shall use the functional form $\mathcal{A}_{\text{closed}}^{\text{dilaton-dilatino}}(\tilde{p})$($\equiv \mathcal{A}_{\text{string}}^{\text{10d}}(\tilde{p})$ of equation (\ref{first-amplitude-10d})) but considering that this is immersed in this curved space. This implies that we consider the dilaton and dilatino wave functions, as well as, the Dirac matrices $\widetilde{\Gamma}$ defined in AdS$_5 \times S^5$, and $\tilde{s} \rightarrow \tilde{s}(r)$. Therefore, we label this modification of the ten-dimensional scattering amplitude with a tilde $\tilde{\mathcal{A}}_{\text{closed}}^{\text{dilaton-dilatino}}(\tilde{p})$, as shown in equation (\ref{A4}). In this equation $g$ is the determinant of the metric of AdS$_5 \times S^5$ background, written as:
\begin{eqnarray}
ds^2 = \frac{r^2}{R^2} \eta_{\mu \nu} dx^\mu dx^\nu + \frac{R^2}{r^2} dr^2 + R^2 d\Omega_5^2 \ ,
\end{eqnarray}
where the radius of AdS$_5$ and $S^5$ is given by the relation  $R^4 = 4 \pi g_s N_c \alpha'^2$, being $g_s$ the string coupling and $\alpha'$ the string constant. The four-dimensional Minkowski spacetime Lorentz indices are denoted by lower case Greek letters $\mu, \nu, \dots $, and the corresponding metric is $\eta_{\mu \nu}$. We should emphasize that one needs to introduce a deformation in the AdS$_5 \times S^5$ spacetime represented by a minimum radius $r_0$ in the radial coordinate. This is related to a certain confinement energy scale $\Lambda$ in the dual gauge theory, $r_0 \sim \Lambda R^2$. In addition, $N_c$ is the rank of the gauge group $SU(N_c)$ of the dual gauge theory, while in the type IIB superstring theory configuration $N_c$ represents the number of coincident D3-branes. In the holographic dual gauge field theory we can define the
string tension as $\hat{\alpha}' \equiv 1/(\lambda_{\text{'t Hooft}}^{1/2} \Lambda^2)$, where $\lambda_{\text{'t Hooft}} = g_{YM}^2 N_c \equiv 4 \pi g_s N_c$ is the 't Hooft coupling.

Now, let us focus on the radial integral (\ref{A4}). Generically speaking for any external string-state wave function, the variation of the function  $e^{i p_\mu x^\mu}$, which corresponds to a plane wave propagating in the four-dimensional Minkowski spacetime, depends on $r$ through $p \sim r/R^2$ and it varies on the string scale. However, the radial- and angular-dependent functions of this ten-dimensional wave function are slow varying. Therefore, the effective four-dimensional scattering amplitude can be obtained by the integration of the  scattering amplitude of four closed strings over  $(r, \Omega_5)$ as shown in equation (\ref{A4}). A key point emphasized in reference \cite{Polchinski:2001tt} is that the four momentum of the dual gauge theory is $p_\mu=-i \partial_\mu$. If an inertial observer is localized at position $r$ in the AdS$_5$ interior, there is the following relation between the ten-dimensional momentum $\tilde{p}^\mu$ and the four-dimensional one $p^\mu$: 
\begin{eqnarray}
\tilde{p}^\mu = \frac{R}{r} p^\mu \ , \label{losp}
\end{eqnarray}
obviously, this only concerns to the directions $\mu=0, 1, 2, 3$.

At this point we can calculate  $\mathcal{A}_{4}^{\text{glueball-fermion}}(p)$. In order to do it we must consider the explicit form of the dilaton and dilatino wave functions, which we study next.

\subsection{Dilaton wave function}

As mentioned we need the explicit form of the dilaton and dilatino wave functions on  AdS$_5 \times S^5$, considering the infrared deformation $r_0$. The dilaton wave function is given by:
\begin{eqnarray}
\Phi_{\Delta_i}(x, r, \Omega_5) = e^{i p_i \cdot x} \varphi_{\Delta_i}(r, \Omega_5) \, . \label{dilaton-10d}
\end{eqnarray}
We are interested in the high-energy behavior of the scattering amplitude. Therefore, we focus on the large-$r$ behavior of this wave function which can be factorized as follows:
\begin{eqnarray}
\varphi_{\Delta_i} (r,\Omega_5) \approx \frac{c_i}{R^4\Lambda} \left(\frac{r_0}{r}\right)^{\Delta_i}  Y_{\Delta_i}^{(0)}(\Omega_5) \, . \label{dilaton-6d}
\end{eqnarray}
The scalar spherical harmonics on $S^5$ are denoted by $Y_{\Delta_i}^{(0)}(\Omega_5)$. In addition, $\Delta_i$ is the conformal dimension of the lowest conformal dimension operator ${\cal {O}}_{k_i}^{(8)}(x_\mu)$ (with $\Delta_i=k_i+4$, $k_i \ge 0$) which creates the corresponding $i$-th glueball state in the dual gauge field theory. The orthonormalization condition for these wave functions is given by:
\begin{eqnarray}
\int_{r_0}^{\infty} dr  \int_{S^5} d \Omega_5 \ \sqrt{g_{\perp}} \ \frac{r^2}{R^2} \ \varphi_{\Delta_i}^*(r,\Omega_5) \ \varphi_{\Delta_j}(r,\Omega_5) = \delta_{\Delta_i \Delta_j} \, ,
\end{eqnarray}
where $\sqrt{g_{\perp}}=\frac{R^6}{r}$. Also, the scalar spherical harmonics satisfy the following condition:
\begin{equation}
\delta_{\Delta_i,\Delta_j} = \int d \Omega_5 \ \sqrt{\hat{g}_{S^5}} \ Y^{(0)*}_{\Delta_i}(\Omega_5) \ Y^{(0)}_{\Delta_j}(\Omega_5) \  ,
\end{equation}
where $\sqrt{\hat{g}_{S^5}}=1$ and $\sqrt{g_{S^5}}=R^5$. The radial integral
\begin{eqnarray}
R^4 \int_{r_0}^\infty dr \ r \ \frac{|c_i|^2}{R^8\Lambda^2}  \left(\frac{r_0}{r}\right)^{2\Delta_i} = 1 \, ,
\end{eqnarray}
allows to calculate the normalization constant $|c_i| = \sqrt{2(\Delta_i-1)}$.

\subsection{Dilatino wave function}

Now, we consider the 32-component Majorana-Weyl right-handed spinor from type IIB superstring theory, i.e. the right-handed dilatino, which has only 16 non-vanishing components, given by:  
\begin{eqnarray}
\lambda_i=\begin{pmatrix}
0 \\ \lambda'_i 
\end{pmatrix} \, . \label{32-spinor}
\end{eqnarray}
From the spontaneous compactification of type IIB supergravity on AdS$_5 \times S^5$, the right-handed dilatinos have the following Kaluza-Klein decomposition:  
\begin{eqnarray}
\lambda'_i(y,\Omega_5)=\tilde{\lambda}_{i}(y)\otimes Y^{(1/2)}_{\tilde{\Delta}_i}(\Omega_5) \, , \label{dilatino-1}
\end{eqnarray}
being $\tilde{\lambda}_{i}(y)$ a four-component spinor on AdS$_5$ with $y = (x^\mu, r)$, while $Y^{(1/2)}_{\tilde{\Delta}_i}(\Omega_5)$ denotes a spinor spherical harmonic on the five-sphere  satisfying the orthonormalization condition:
\begin{equation}
\delta_{\tilde{\Delta}_i,\tilde{\Delta}_j} = \int d \Omega_5 \ \sqrt{\hat{g}_{S^5}} \ Y^{(1/2)\dagger}_{\tilde{\Delta}_i}(\Omega_5) \ Y^{(1/2)}_{\tilde{\Delta}_j}(\Omega_5) \  .
\end{equation}
These fermionic Kaluza-Klein fields are dual to the $\mathcal{O}^{(6)}_k(x)$ operators of $\mathcal{N}=4$ SYM theory that we shall discuss in section 5. The general solution for $\lambda'_i(y,\Omega_5)$ is factorizable as the product of a function of the $y$-variables of the form $e^{i p \cdot x} f'_{i}(r/r_0)$, multiplied by the corresponding spinor spherical harmonic $Y^{(1/2)}_{\tilde{\Delta}_i}(\Omega_5)$.
%
%
%
%
The five-dimensional dilatino $\tilde{\lambda}_{i}(y)$ in  (\ref{dilatino-1}) is a solution of the Dirac's equation on AdS$_5$:
\begin{eqnarray}
\left(z\gamma^m\partial_m-2\gamma^5+m_k\right)\tilde{\lambda}_{i}(y)=0 \, , \label{Dirac-5d}
\end{eqnarray}
with the Kaluza-Klein mass $m_k=k+\frac{3}{2}$, being $k \ge 0$. Equation (\ref{Dirac-5d}) has a general solution in terms of the $z$ variable given by:
\begin{eqnarray}
\tilde{\lambda}_{i}(y)= C_i e^{i p \cdot x} z^{5/2}\left[ a^+(p)J_{m R-1/2}(M z) + a^-(p)J_{m R+1/2}(M z)\right] \, , \label{5-d-spinor-solution}
\end{eqnarray}
where we have changed variables according to $z=R^2/r$. Also,
we have written this solution in terms of Bessel functions and five-dimensional Dirac spinors $a_{\pm}$, from which one can construct four-dimensional Dirac spinors $v_{\sigma}$. They satisfy the Dirac's equation in four dimensions, namely: $i\gamma^{\mu}p_{\mu} \ v_{\sigma}=M v_{\sigma}$, and also the four-dimensional relativistic dispersion relation: $p^\mu p_\mu=-M^2$. There is a relation between the five-dimensional Dirac spinors $a_\pm$ as follows:
\begin{eqnarray}
a_+&  = & \frac{i\gamma^{\mu} p_{\mu}}{M}a_- \, .
\end{eqnarray}
Then, the solution (\ref{5-d-spinor-solution}) becomes:
\begin{eqnarray}
\tilde{\lambda}_{i}(y)= C_i e^{i p \cdot x} z^{5/2}\left[ P_+J_{mR-1/2}(Mz) + P_-J_{mR+1/2}(Mz)\right] v_{i} \, , \label{dilatino-sol-2}
\end{eqnarray}
where $P_{\pm}=\frac{(1 \pm \gamma^5)}{2}$ are the chiral projection operators. Recall that we need to consider the 
scattering amplitude at high energy, i.e. for large $r$. This implies that $p \sim\sqrt{s} \sim\frac{r}{R^2}$ must be large and the Bessel functions $J(Mz)\sim J(M/\sqrt{s})$ can be approximated by: 
\begin{eqnarray}
J_{m R \pm 1/2}(M z) \sim \frac{(M z)^{m R \pm 1/2}}{2^{m R \pm 1/2}(m R \pm 1/2)!} \, .
\end{eqnarray}
Therefore, we have:
\begin{eqnarray}
\tilde{\lambda}_{i}(y)&\approx& e^{i p \cdot x}\frac{\tilde{c}_i}{R^{9/2}\Lambda^{3/2}(\tilde{\Delta}_i-5/2)!}(r/r_0)^{-\tilde{\Delta}_i}\left[ P_{+} + \frac{M_iR^2}{2r(\tilde{\Delta}_i-3/2)}P_-\right]v_{i} \, , \label{dilatino-large-r}
\end{eqnarray}
where $\tilde{c}_i$ is obtained from the normalization condition, leading to:
\begin{eqnarray}
\tilde{c}_i=\sqrt{2 \Lambda R(\tilde{\Delta}_i-1)} \ (\tilde{\Delta}_i-5/2)! \, .
\end{eqnarray}

\subsection{Glueball-fermion scattering at fixed angle}

Below we have obtained the dilaton and dilatino wave functions needed to calculate the four-dimensional glueball-fermion scattering amplitude. Now, we use the equations developed in section 2 in order to calculate $\mathcal{A}_{4}^{\text{glueball-fermion}}(p)$ from equation (\ref{A4}). 
The idea is the following one. One has to start from the dilaton-dilatino scattering amplitude in ten-dimensional Minkowski spacetime obtained from type IIB superstring theory, and following Polchinski-Strassler construction, consider a local approximation in the AdS$_5 \times S^5$ bulk.

Taking into account the relations between the Mandelstam variables and the scattering angle (\ref{kinematics})
equation (\ref{A4}) becomes:
\begin{eqnarray}
&&\mathcal{A}_{4}^{\text{glueball-fermion}}(p) = \frac{4\pi g_s^2\alpha'^4(49 \cos (3 \theta)+266 \cos (2 \theta)-97 \cos (\theta)+550)}{65536} \times \nonumber \\
&& \,\,\,\,\,\,\,\,\,\,\,\,\,\,\,\,\,\,\,\,\,\,\, \int_{r_0}^{\infty} dr \int_{S^5} d\Omega_5  \ r^3 R^2 \ (\alpha'\tilde{s})^3 \bar{\lambda}_1(y,\Omega_5)\Phi_2(y,\Omega_5)\Phi_3(y,\Omega_5)(g_{MN}k_3^M \widetilde{\Gamma}^N)\lambda_4(y,\Omega_5) \nonumber\\
&& \,\,\,\,\,\,\,\,\,\,\,\,\,\,\,\,\,\,\,\,\,\,\, 
\times \prod_{\chi=\tilde{s},\tilde{t},\tilde{u}}\frac{\Gamma(-\alpha'\chi/4)}{\Gamma(1-\alpha'\chi/4)} \, . \label{amplitude-20}
\end{eqnarray}
Note that $\lambda_1(y,\Omega_5)$ and $\lambda_4(y,\Omega_5)$ are 32-component Majorana-Weyl right-handed spinors of type IIB supergravity from equation (\ref{32-spinor}). $\widetilde{\Gamma}_a$ is a Dirac gamma matrix in AdS$_5 \times S^5$ written in the following representation:
\begin{eqnarray}
\widetilde{\Gamma}^a= \sigma^1 \otimes 1_4 \otimes \widetilde{\gamma}^a \, , \label{Gamma-curved}
\end{eqnarray}
being $\widetilde{\gamma}^a$ a Dirac gamma matrix where $a=0, \dots, 4$ represent curved-space indices in AdS$_5$. The relation between the AdS$_5$ Dirac gamma matrices $\widetilde{\gamma}^a$ and flat-space Dirac gamma matrices $\gamma^{\hat{a}}$, where $\hat{a}=0, \dots, 4$ are flat-space indices, is given by:
\begin{equation}
    \widetilde{\gamma}_a=\frac{r}{R}\gamma_{\hat{a}} \, .
\end{equation}
We consider the following representation for the Dirac gamma matrices in four-dimensional Minkowski spacetime:
\begin{eqnarray}
\gamma^{\mu}=\begin{pmatrix} 
0 &-i\sigma^{\mu} \\ -i\bar{\sigma}^{\mu} & 0
    \end{pmatrix} \, ,
\end{eqnarray}
which are written in terms of the Pauli matrices:
\begin{eqnarray}
\sigma^1=\begin{pmatrix}0 & 1 \\ 1 & 0 
\end{pmatrix}, \,\,\,\,\,\,\,\,\,\,\,
\sigma^2=\begin{pmatrix}0 & -i \\ i & 0
\end{pmatrix}, \,\,\,\,\,\,\,\,\,\,\,
\sigma^3=\begin{pmatrix}1 & 0\\ 0& -1
\end{pmatrix} \, .
\end{eqnarray}
Let us focus on the factor:
\begin{equation}
\bar{\lambda}_1(y,\Omega_5) \ \Phi_2(y,\Omega_5) \ \Phi_3(y,\Omega_5) \ (g_{MN}k_3^M \widetilde{\Gamma}^N) \ \lambda_4(y,\Omega_5) \label{integrand-1} \, .
\end{equation}
Considering equations (\ref{dilaton-10d}), (\ref{dilaton-6d}), (\ref{32-spinor}), (\ref{dilatino-1}) and (\ref{Gamma-curved}), it can be rewritten as:
\begin{eqnarray}
&&  \left(0, \tilde{\lambda}^\dagger_{1}(y)\otimes Y^{(1/2)\dagger}_{\tilde{\Delta}_1}(\Omega_5)\right) \widetilde{\Gamma}^0  \ \widetilde{\Gamma}^N \begin{pmatrix}
0 \\ \tilde{\lambda}_{4}(y)\otimes Y^{(1/2)}_{\tilde{\Delta}_4}(\Omega_5) 
\end{pmatrix} = \nonumber \\
&& \left(0, \tilde{\lambda}^\dagger_{1}(y)\otimes Y^{(1/2)\dagger}_{\tilde{\Delta}_1}(\Omega_5)\right) \begin{pmatrix}0 & 1_4 \otimes \widetilde{\gamma}^0 \\  1_4 \otimes \widetilde{\gamma}^0 & 0 
\end{pmatrix}   \begin{pmatrix}0 & 1_4 \otimes \widetilde{\gamma}^b \\  1_4 \otimes \widetilde{\gamma}^b & 0 
\end{pmatrix} \begin{pmatrix}
0 \\ \tilde{\lambda}_{4}(y)\otimes Y^{(1/2)}_{\tilde{\Delta}_4}(\Omega_5) 
\end{pmatrix}, \nonumber \\
&&
\end{eqnarray}
which leads to:
\begin{eqnarray}
&& \tilde{\lambda}^\dagger_{1}(y) \ \widetilde{\gamma}^0   \ \widetilde{\gamma}^b \ \tilde{\lambda}_{4}(y) \otimes Y^{(1/2)\dagger}_{\tilde{\Delta}_1}(\Omega_5) \ Y^{(1/2)}_{\tilde{\Delta}_4}(\Omega_5)  \, ,
\end{eqnarray}
where the first factor $ \tilde{\lambda}^\dagger_{1}(y) \ \widetilde{\gamma}^0   \ \widetilde{\gamma}^b \ \tilde{\lambda}_{4}(y)$ is defined on AdS$_5$ while the second factor is on $S^5$. We may assume the ten-dimensional kinematics of a head-on collision of closed strings given by:
\begin{eqnarray}
    k_1^{M}&=& \left( \frac{\sqrt{\tilde{s}}}{2},\frac{\sqrt{\tilde{s}}}{2}, 0, \dots, 0 \right) \, , \nonumber\\
    k_2^{M}&=& \left(\frac{\sqrt{\tilde{s}}}{2},-\frac{\sqrt{\tilde{s}}}{2}, 0, \dots, 0 \right) \, , \nonumber\\
    k_3^{M}&=& \left(-\frac{\sqrt{\tilde{s}}}{2},\frac{\sqrt{\tilde{s}}}{2}\cos\theta,\frac{\sqrt{\tilde{s}}}{2}\sin\theta, 0, \dots, 0 \right) \, , \nonumber\\
    k_4^{M}&=& \left(-\frac{\sqrt{\tilde{s}}}{2},-\frac{\sqrt{\tilde{s}}}{2}\cos\theta,-\frac{\sqrt{\tilde{s}}}{2}\sin\theta, 0, \dots, 0 \right) \, .
\end{eqnarray}
This allows to consider that the Lorentz index $N$ now reduces to $b=0, 1, 2$. Then, the factor (\ref{integrand-1}) leads to the angular integral:
\begin{equation}
I_{\tilde{\Delta}_1,\Delta_2,\Delta_3,\tilde{\Delta}_4} = \int_{S^5} d\Omega_5 \  \sqrt{\hat{g}_{S^5}}   \ Y^{(0)^*}_{{\Delta}_2}(\Omega_5) \ Y^{(0)}_{{\Delta}_3}(\Omega_5) \  Y^{(1/2)\dagger}_{\tilde{\Delta}_1}(\Omega_5) \ Y^{(1/2)}_{\tilde{\Delta}_4}(\Omega_5)  \, . \label{angular-integral}
\end{equation}
In four dimensions we may define the momenta of the four particles which participate in the scattering process as:
\begin{eqnarray}
    p_1^{\mu}&=&\left(\frac{\sqrt{s}}{2}, \sqrt{\frac{s}{4}-M^2}, 0, 0 \right) \, , \nonumber\\
    p_2^{\mu}&=&\left(\frac{\sqrt{s}}{2}, -\sqrt{\frac{s}{4}-M^2}, 0, 0 \right) \, , \nonumber\\
    p_3^{\mu}&=&\left(-\frac{\sqrt{s}}{2}, \sqrt{\frac{s}{4}-M^2} \cos\theta, \sqrt{\frac{s}{4}-M^2} \sin\theta, 0 \right) \, , \nonumber\\
    p_4^{\mu}&=&\left(-\frac{\sqrt{s}}{2}, -\sqrt{\frac{s}{4}-M^2} \cos\theta, -\sqrt{\frac{s}{4}-M^2} \sin\theta, 0 \right) \, , \label{K4d}
\end{eqnarray}
being particles 2 and 3 glueballs of mass $M_{\text{glueball}} \simeq 1$ GeV, while particles 1 and 4 are spin-1/2 fermions with mass $M_{\text{fermion}} \simeq 1$ GeV. Thus, in what follows we shall assume that $M_{\text{glueball}} \simeq M_{\text{fermion}} \simeq M \simeq 1$ GeV. This leads to:\footnote{We use the wave functions of the dilaton and the dilatino for large values of $r$ given by equations (\ref{dilaton-6d}) and (\ref{dilatino-large-r}), respectively. For this reason the result is an approximation as indicated in equation (\ref{amplitude-31}). Importantly, this does no affect the $s$-power behavior of any of these expressions.}
\begin{eqnarray}
&&\mathcal{A}_{4}^{\text{glueball-fermion}}(p)  \approx \frac{2\pi g_s^2\alpha'^4(49 \cos (3 \theta)+266 \cos (2 \theta)-97 \cos (\theta)+550)}{65536}   \nonumber \\
&& \,\,\,\,\,\,\,\,\,\,\,\,\,\,\,\,\,\,\,\,\,\,\,\,\,\,\, \times \ I_{\tilde{\Delta}_1,\Delta_2,\Delta_3,\tilde{\Delta}_4} \ \int_{r_0}^{\infty} dr \ r^4 R  \ (\alpha'\tilde{s})^3 \sqrt{\tilde{s}} \  \frac{c_2^* c_3}{R^8 \Lambda^2} \left(\frac{r_0}{r}\right)^{\Delta_2+\Delta_3}  \nonumber \\
&& \,\,\,\,\,\,\,\,\,\,\,\,\,\,\,\,\,\,\,\,\,\,\,\,\,\,\, \times \ \bar{\tilde{\lambda}}_1(y)(\gamma^0+\cos\theta \gamma^1+\sin\theta\gamma^2)\tilde{\lambda}_4(y)\prod_{\chi=\tilde{s},\tilde{t},\tilde{u}}\frac{\Gamma(-\alpha'\chi/4)}{\Gamma(1-\alpha'\chi/4)} \, . \label{amplitude-31}
\end{eqnarray}
Recall that the four-dimensional Dirac's spinor is defined as:
\begin{eqnarray}
v_i = \begin{pmatrix}
\sqrt{p_i \cdot \sigma} \ \zeta^s \\ \sqrt{p_i \cdot \bar{\sigma}} \ \zeta^s
\end{pmatrix} \, ,
\end{eqnarray}
where $s=1, 2$, with: 
\begin{eqnarray}
\zeta^1 = \begin{pmatrix}
1 \\ 0
\end{pmatrix} \,\,\,\,\,\,\,\,  \text{and} \,\,\,\,\,\,\,\,  \zeta^2 = \begin{pmatrix}
0 \\ 1
\end{pmatrix} \ .
\end{eqnarray}
Now, for fermions 1 and 4 we shall use the following expressions:
\begin{eqnarray}
    \sqrt{p\cdot \sigma}=\frac{p\cdot \sigma +M}{\sqrt{2(p^0+M)}} \, ,
\end{eqnarray}
and the corresponding one with $\bar {\sigma}$, which considering the kinematics (\ref{K4d}) give:
\begin{eqnarray}
\sqrt{p_1\cdot \sigma}=\left(\frac{1}{\sqrt{s}+2M)}\right)^{1/2}\begin{pmatrix}
  -\frac{\sqrt{s}}{2}+M & \sqrt{\frac{s}{4}-M^2} \\ \sqrt{\frac{s}{4}-M^2} & -\frac{\sqrt{s}}{2}+M  
 \end{pmatrix}     \, ,
\end{eqnarray}
and
\begin{eqnarray}
\sqrt{p_1\cdot \bar{\sigma}}=\left(\frac{1}{\sqrt{s}+2M)}\right)^{1/2}\begin{pmatrix}
  -\frac{\sqrt{s}}{2}+M & -\sqrt{\frac{s}{4}-M^2} \\ -\sqrt{\frac{s}{4}-M^2} & -\frac{\sqrt{s}}{2}+M  
 \end{pmatrix}     \, ,
\end{eqnarray}
for the first fermion, and
\begin{eqnarray}
    \sqrt{p_4 \cdot \sigma}=\left(\frac{1}{-\sqrt{s}+2M}\right)^{1/2}\begin{pmatrix}
        \frac{\sqrt{s}}{2}+M & -\sqrt{\frac{s}{4}-M^2}e^{-i\theta} \\ -\sqrt{\frac{s}{4}-M^2}e^{i\theta} & \frac{\sqrt{s}}{2}+M   
    \end{pmatrix}  \, ,  
\end{eqnarray}
and
\begin{eqnarray}
    \sqrt{p_4 \cdot \bar{\sigma}}=\left(\frac{1}{-\sqrt{s}+2M}\right)^{1/2}\begin{pmatrix}
        \frac{\sqrt{s}}{2}+M & \sqrt{\frac{s}{4}-M^2}e^{-i\theta} \\ \sqrt{\frac{s}{4}-M^2}e^{i\theta} & \frac{\sqrt{s}}{2}+M   
    \end{pmatrix}    \, ,
\end{eqnarray}
for the fourth fermion.

Focusing on the high-energy limit at fixed angle, the three ten-dimensional Mandelstam variables are parametrically large. Thus, we may use the Stirling's formula for the Euler's gamma functions and approximate the following expression as follows:
\begin{eqnarray}
&&\frac{\Gamma(-\frac{\alpha'\tilde{s}}{4})\Gamma(-\frac{\alpha'\tilde{u}}{4})\Gamma(-\frac{\alpha'\tilde{t}}{4})}{\Gamma(1+\frac{\alpha'\tilde{s}}{4})\Gamma(1+\frac{\alpha'\tilde{u}}{4})\Gamma(1+\frac{\alpha'\tilde{t}}{4})}
= \nonumber \\
&&(-1)^{1+\alpha'\tilde{s}/2} \left(\frac{128e^2}{\alpha'^3}\right)  \frac{\sin(\frac{\pi \alpha'\tilde{t}}{4})\sin(\frac{\pi \alpha'\tilde{u}}{4})}{\sin(\frac{\pi \alpha'\tilde{s}}{4})}  \frac{(\alpha'\tilde{s}/4)^{-\alpha'\tilde{s}/2}(\alpha'\tilde{u}/4)^{-\alpha'\tilde{u}/2}(\alpha'\tilde{t}/4)^{-\alpha'\tilde{t}/2}}{\tilde{s}\tilde{t}\tilde{u}} \, . \nonumber \\
&&
\end{eqnarray}
We can further work out this expression using the scattering angle through equations (\ref{kinematics}) together with the dispersion relation $\tilde{s}+\tilde{t}+\tilde{u}=0$, also including the dilaton and dilatino wave functions that we have obtained in the previous subsections 3.2 and 3.3 respectively.
Thus, we obtain:
\begin{eqnarray}
&&\mathcal{A}_{4}^{\text{glueball-fermion}}(p) = \frac{512 \ i  \ e^2 \pi  g_s^2\alpha'^4(\Delta-2)(\tilde{\Delta}-2)}{65536 \ \Lambda^{4-(\Delta+\tilde{\Delta})} \ R^{14-2(\Delta+\tilde{\Delta})}} \ I_{\tilde{\Delta}_1,\Delta_2,\Delta_3,\tilde{\Delta}_4} \times \nonumber \\
&&\frac{(49 \cos (3 \theta)+266 \cos (2 \theta)-97 \cos (\theta)+550)}{\sin^2\theta}  \int_{r_0}^{\infty} dr  e^{-2\beta_{\tilde{s}\tilde{t}\tilde{u}}}r^{3-(\Delta+\tilde{\Delta})}\sqrt{s}\left(1+\frac{M^2R^4}{r^2(\tilde{\Delta}-3)^2}\right) \nonumber \\
&& \times \Bigg[\sqrt{s-4M^2}(1+\cos\theta)+\cos\theta[\sqrt{s}(1+\cos\theta)+2M(1-\cos\theta)]
+\sin^2\theta(\sqrt{s}-2M)\Bigg] \, , \label{A1-1} \nonumber \\
&&
\end{eqnarray}
where we have defined:
\begin{equation}
\beta_{\tilde{s}\tilde{t}\tilde{u}}=\frac{\alpha'\tilde{s}}{4}\log\left(\frac{\alpha'\tilde{s}}{4}\right)+\frac{\alpha'\tilde{t}}{4}\log\left(\frac{\alpha'\tilde{t}}{4}\right)+\frac{\alpha'\tilde{u}}{4}\log\left(\frac{\alpha'\tilde{u}}{4}\right)    \, ,
\end{equation}
or equivalently:
\begin{equation}
 \beta_{\tilde{s}\tilde{t}\tilde{u}}=-\frac{\alpha'\tilde{s}}{4}\left( i\pi+\frac{1}{2}(1-\cos\theta)\log\left(\frac{1-\cos\theta}{2}\right)+\frac{1}{2}(1+\cos\theta)\log\left(\frac{1+\cos\theta}{2}\right)\right) \, ,
\end{equation}
where assuming elastic scattering we set $\Delta=2\Delta_i$ and $\tilde{\Delta}=2 \tilde{\Delta}_i$. Taking into account that from the metric warp factor the ten-dimensional and the four-dimensional Mandelstam variables are related by $\tilde{s}=R^2 s/r^2$. Then, changing variables as $\rho=\frac{r}{r_0}$ with $r_0=\Lambda R^2$, and setting $\Lambda\sim M$, equation (\ref{A1-1}) becomes:
\begin{eqnarray}
&&\mathcal{A}_{4}^{\text{glueball-fermion}}(p) = 
\frac{512 \ i \ e^2 \pi g_s^2 \alpha'^4 (\Delta-2)(\tilde{\Delta}-2)}{65536 \ R^6} \times \nonumber\\
&& \,\,\,\,\,\,\,\,\,\,\,\, \frac{(49 \cos (3 \theta)+266 \cos (2 \theta)-97 \cos (\theta)+550)}{\sin^2\theta} \nonumber\\
&& \times \int_{1}^{\infty} d\rho e^{\frac{\alpha's}{2\Lambda^2R^2\rho^2}\left( i\pi+\frac{1}{2}(1-\cos\theta)\log(\frac{1-\cos\theta}{2})+\frac{1}{2}(1+\cos\theta)\log(\frac{1+\cos\theta}{2})\right)}\rho^{3-(\Delta+\tilde{\Delta})}\sqrt{s}\left(1+\frac{1}{\rho^2(\tilde{\Delta}-3)^2}\right)\nonumber\\
&& \times \Bigg[\sqrt{s-4\Lambda^2}(1+\cos\theta)+\cos\theta[\sqrt{s}(1+\cos\theta)+2\Lambda(1-\cos\theta)]\nonumber\\
&& + \sin^2\theta(\sqrt{s}-2\Lambda)\Bigg]I_{\tilde{\Delta}_1,\Delta_2,\Delta_3,\tilde{\Delta}_4} \, .
\end{eqnarray}
The radial integral in the variable $\rho$ can be solved analytically leading to:
\begin{eqnarray}
&&\mathcal{A}_{4}^{\text{glueball-fermion}}(p) =\frac{i\ e^2 \pi g_s^2 \alpha'^{5-\frac{\Delta+\tilde{\Delta}}{2}} (\Delta-2)(\tilde{\Delta}-2)}{128 R^{6}(\tilde{\Delta}-3)^2}  \nonumber \\
&& \times \ \frac{(49 \cos (3 \theta)+266 \cos (2 \theta)-97 \cos (\theta)+550)}{\sin^2\theta} \ 2^{\Delta+\tilde{\Delta}-5} \ \Bigg(-\frac{\Lambda^2 R^2}{ f(\theta)}\Bigg)^{\frac{1}{2} (\Delta+\tilde{\Delta}-2)} \nonumber \\
&& \times \ \Bigg\{\frac{1}{\Lambda^2 R^2}\Gamma
   \left[\frac{1}{2} (\Delta+\tilde{\Delta}-4)\right] \Bigg(-\alpha' (\tilde{\Delta}-3)^2 f(\theta)   +2 \frac{\Lambda^2 R^2}{s} (\Delta+\tilde{\Delta}-4)\Bigg) \nonumber\\
&& + \frac{\alpha'(\tilde{\Delta}-3)^2 }{\Lambda^2 R^2} \ f(\theta) \ \Gamma \Bigg[\frac{1}{2} (\Delta+\tilde{\Delta}-4),\frac{-\alpha' s f(\theta)}{4 \Lambda^2 R^2}\Bigg]-\frac{4}{s} \ \Gamma \Bigg[\frac{1}{2} (\Delta+\tilde{\Delta}-2),\frac{-\alpha' s f(\theta)}{4 \Lambda^2 R^2}\Bigg]\Bigg\} \nonumber\\
&& \times \Big[\left(s^{3-\frac{\Delta+\tilde{\Delta}}{2}}-2\Lambda^2s^{2-\frac{\Delta+\tilde{\Delta}}{2}}-2\Lambda^4s^{1-\frac{\Delta+\tilde{\Delta}}{2}}-...\right)(1+\cos\theta)+\cos\theta[s^{3-\frac{\Delta+\tilde{\Delta}}{2}}(1+\cos\theta) \nonumber\\
&& + 2 \Lambda s^{5/2-\frac{\Delta+\tilde{\Delta}}{2}}(1-\cos\theta)] + \sin^2\theta(s^{3-\frac{\Delta+\tilde{\Delta}}{2}}-2\Lambda s^{5/2-\frac{\Delta+\tilde{\Delta}}{2}})\Big] \ I_{\tilde{\Delta}_1,\Delta_2,\Delta_3,\tilde{\Delta}_4} \, , \label{A4-integrated}
\end{eqnarray}
where:
\begin{equation}
f(\theta)= (1+\cos (\theta )) \log \left(\cos ^2\left(\frac{\theta }{2}\right)\right)+(1-\cos (\theta )) \log
\left(\sin ^2\left(\frac{\theta }{2}\right)\right)+2 i \pi \, .
\end{equation}
In equation (\ref{A4-integrated}) $\Gamma[z]$ stands for the Euler's gamma function, while $\Gamma[z, x]$ indicates the incomplete gamma function.
In the derivation of equation (\ref{A4-integrated}) we have used the Taylor expansion for $s \gg 4 \Lambda^2$, leading to:
\begin{eqnarray}
    \sqrt{s-4\Lambda^2}=\sqrt{s}\sqrt{1-\frac{4\Lambda^2}{s}}\sim\sqrt{s}\left(1-\frac{2\Lambda^2}{s}-\frac{2\Lambda^4}{s^2}- \dots \right) \, . \label{Amplitude-4}
\end{eqnarray}
At this point it is worth emphasizing that the above four-dimensional scattering amplitude can be rewritten in terms of the twists of the operators which create the confining gauge theory states in the following way:
\begin{eqnarray}
&&\mathcal{A}_{4}^{\text{glueball-fermion}}(p) =  \frac{i \ e^2 \pi g_s^2 \alpha'^{\frac{9-{\mathcal{T}}}{2}}(\tau-2)(\tilde{\tau}-1)}{64 \ R^{6} \ (\tilde{\tau}-2)^2} \times \nonumber\\
&& \frac{(49 \cos (3 \theta)+266 \cos (2 \theta)-97 \cos (\theta)+550)}{\sin^2\theta} \ 2^{{\mathcal{T}}-4} \Bigg(-\frac{\Lambda^2 R^2}{ f(\theta)}\Bigg)^{\frac{1}{2} ({\mathcal{T}}-1)} \times \nonumber \\
&& \Bigg\{\frac{1}{\Lambda^2 R^2}\Gamma
   \left[\frac{1}{2} ({\mathcal{T}}-3)\right] \Bigg(-\alpha' (\tilde{\tau}-2)^2 f(\theta) + 2 \frac{\Lambda^2 R^2}{s} ({\mathcal{T}}-3)\Bigg) + \nonumber\\
&&\frac{\alpha'(\tilde{\tau}-2)^2 }{\Lambda^2 R^2}f(\theta)\Gamma \Bigg[\frac{1}{2} ({\mathcal{T}}-3),\frac{-\alpha' s f(\theta)}{4 \Lambda^2 R^2}\Bigg]-\frac{4}{s} \Gamma \Bigg[\frac{1}{2} ({\mathcal{T}}-1),\frac{-\alpha' s f(\theta)}{4 \Lambda^2 R^2}\Bigg]\Bigg\} \times \nonumber\\
&& \Big[s^{5/2-\mathcal{T}/2}(1+\cos\theta)+\Lambda s^{2-\mathcal{T}/2}(\cos\theta-1)-\Lambda^2s^{3/2-\mathcal{T}/2}-\Lambda^4s^{1/2-\mathcal{T}/2}+...\Big]I_{\tilde{\tau}_1,\tau_2,\tau_3,\tilde{\tau}_4} \, . \label{A4final} \nonumber \\ 
&&
\end{eqnarray}
where we have expressed the powers of $s$ in terms of the twists corresponding to $\tau=\Delta = 2 \Delta_i$ and $\tilde{\tau}=\tilde{\Delta}-1=2 (\tilde{\Delta}_i-\frac{1}{2})$, being the total twist $\mathcal{T}=\tau+\tilde{\tau}$.  This behavior is expected for confining quantum field theories from the analysis of the scaling laws for large-momentum-transfer exclusive processes at fixed angle with independent interactions among the hadron constituents \cite{Brodsky:1974vy}, which leads to:
\begin{eqnarray}
\mathcal{A} \propto s^{2-n/2} \ s^{L/2-1/2}    \, , \label{EqBrodskyFarrar}
\end{eqnarray}
where $L$ is the number of pairs of constituents from different hadrons which have a large-angle scale-invariant interaction, while $n \equiv \mathcal{T}$ is the total twist of the hadrons involved in the process. Comparing equation (\ref{EqBrodskyFarrar}) with  (\ref{A4final}) we obtain exactly the leading high-energy contribution from equation (\ref{A4final}) with $L=2$:
\begin{eqnarray}
\mathcal{A} \propto s^{5/2-\mathcal{T}/2} \, . \label{EqBrodskyFarrarcomparison}
\end{eqnarray} 
Although, in principle, in our calculation we do not count partons since we work in the strongly coupled regime of ${\cal {N}}=4$ SYM theory with an IR cut-off, this results might be understood as if string theory in certain curved backgrounds in the context of the gauge/string theory duality could reproduce the scaling laws mentioned in the preceding paragraphs.

It is also very interesting to consider the ${\cal {N}}=4$ SYM theory spin-1/2 fermionic operators of the type ${\cal {O}}^{(6)}_k(x)$ and  glueball operators of the form ${\cal {O}}^{(8)}_k(x)$ with the smallest scaling dimensions, i.e. $k=0$, leading to $\Delta^{(8)}=4$ and $\tilde{\Delta}^{(6)}=7/2$, respectively. Thus, $\mathcal{T}=\tau+\tilde{\tau}=8+6=14$, which implies that the differential cross-section for this elastic high-energy process at fixed angle is:
\begin{eqnarray}
\frac{d\sigma}{dt} \propto s^{-11}f(|t|/s) \, , \label{sigma-11}
\end{eqnarray}
which is a prediction of the described calculation.

From equation (\ref{A4final}) we observe an expansion in decreasing powers of $s$ and increasing powers of the confinement energy scale $\Lambda$:
\begin{eqnarray}
&& s^{5/2-\mathcal{T}/2}(1+\cos\theta)+\Lambda s^{2-\mathcal{T}/2}(\cos\theta-1)-\Lambda^2s^{3/2-\mathcal{T}/2}-\Lambda^4s^{1/2-\mathcal{T}/2}+... \, . \label{Lambda-expansion}
\end{eqnarray}
Since $\Lambda \equiv 1/(\hat{\alpha}'^{1/2} \ \lambda_{\text{'t Hooft}}^{1/4})$ and $1 \ll \lambda_{\text{'t Hooft}} \ll N_c$ (recall that $\hat{\alpha}'$ is the string tension of the dual gauge theory), the expansion (\ref{Lambda-expansion}) becomes:
\begin{eqnarray}
&& s^{5/2-\mathcal{T}/2}(1+\cos\theta)+\frac{1}{\hat{\alpha}'^{1/2} \ \lambda_{\text{'t Hooft}}^{1/4}} s^{2-\mathcal{T}/2}(\cos\theta-1)-\frac{1}{\hat{\alpha}' \ \lambda_{\text{'t Hooft}}^{1/2}} s^{3/2-\mathcal{T}/2} \nonumber \\
&& -\frac{1}{\hat{\alpha}'^{2} \ \lambda_{\text{'t Hooft}}} s^{1/2-\mathcal{T}/2}+... \, , \label{tHooft-expansion}
\end{eqnarray}
which turns out to be a strong-coupling expansion resulting from the dual string theory calculation of the gauge theory scattering amplitude at fixed scattering angle. Another interesting remark emergent from the previous equations is the fact that the term proportional to $s^{5/2-\mathcal{T}/2}$ dominates the expansion with the exception of a backward scattering process. Indeed, when the scattering angle approaches $\pi$ the term proportional to $s^{2-\mathcal{T}/2}$ dominates the strong coupling expansion. This is an effect that we can entirely attribute to the strong coupling structure emergent from the string theory dual calculation, which is not manifested in the naive power counting plus the assumptions made about the gauge theory interactions carried out in the calculation of a generic fixed-angle scattering amplitude of references \cite{Brodsky:1973kr,Brodsky:1974vy}.

\section{The Regge limit of the glueball-fermion scattering amplitude}

Now, we investigate the Regge limit $\tilde{s} \gg |\tilde{t}|$ of the glueball-fermion scattering amplitude. From reference \cite{Martin:2025ije} the Regge limit of the kinematic factor corresponding to the scattering of two dilatons and two dilatinos in type II superstring theory in ten-dimensional Minkowski spacetime is:
\begin{eqnarray}
    K_{\text{closed-Regge limit}}^{\text{dilaton-dilatino}}(\tilde{1},2,3,\tilde{4})&=&-\frac{3\tilde{s}^3}{256}\bar{\lambda}_1\Phi_2\Phi_3(k_3\cdot \Gamma)\lambda_4 \, . 
\end{eqnarray}
In order to transform this flat ten-dimensional kinematic factor into the curved-space one, we consider the following ansatz:
\begin{eqnarray}
&&   K_{\text{closed-Regge limit}}^{\text{dilaton-dilatino}}(\tilde{1},2,3,\tilde{4}) \rightarrow   \widetilde{K}_{\text{closed-Regge limit}}^{\text{dilaton-dilatino}}(\tilde{1},2,3,\tilde{4}) \nonumber \\
   &&=-\frac{3\tilde{s}^3}{256}\bar{\lambda}_1(y,\Omega_5) \ \Phi_2(y,\Omega_5) \ \Phi_3(y,\Omega_5) \ (g_{MN}k_3^M  \widetilde{\Gamma}^N) \ \lambda_4(y,\Omega_5)      \, , 
\end{eqnarray}
where $y=(r,x^{\mu})$ while $\widetilde{\Gamma}$ are the Dirac gamma matrices in AdS$_5\times S^5$ defined before. Thus, the Regge limit of the glueball-fermion scattering is given by the following expression:
\begin{eqnarray}
    &&\mathcal{A}_{\text{4-Regge limit}}^{\text{glueball-fermion}}(s,t)=\frac{3\pi g_s^2\alpha'^4}{512} \ I_{\tilde{\Delta}_1,\Delta_2,\Delta_3,\tilde{\Delta}_4}\ \int_{r_0}^{\infty} dr \ r^4 R \ (\alpha'\tilde{s})^3 \ \sqrt{\tilde{s}} \   \frac{c_2^* c_3}{R^8 \Lambda^2} \left(\frac{r_0}{r}\right)^{\Delta_2+\Delta_3}  \nonumber \\
&& \,\,\,\,\,\,\,\,\,\, \times    \ \bar{\tilde{\lambda}}_1(y) \Big(\gamma^0+\gamma^1\Big) \tilde{\lambda_4}(y) \ \frac{\sin[\frac{\pi\alpha'}{4}(\tilde{s}+\tilde{t})]}{\sin(\pi\alpha'\tilde{s}/4)}\left(\frac{\alpha'\tilde{s}}{4}\right)^{-2+\alpha'\tilde{t}/2}e^{2-\alpha'\tilde{t}/2}\frac{\Gamma(-\alpha'\tilde{t}/4)}{\Gamma(1+\alpha'\tilde{t}/4)} \, ,
\end{eqnarray}
which after using the corresponding dilatinos and dilatons wave functions for large $r$ becomes:
\begin{eqnarray}
  && \mathcal{A}_{\text{4-Regge limit}}^{\text{glueball-fermion}}(s,t) \approx \frac{3i\pi g_s^2\alpha'^4(\Delta-2)(\tilde{\Delta}-2)}{512\Lambda^{4-(\Delta+\tilde{\Delta})}R^{14-2(\Delta+\tilde{\Delta})}}I_{\tilde{\Delta}_1,\Delta_2,\Delta_3,\tilde{\Delta}_4}\int_{r_0}^{\infty} dr \ 4^{-\alpha'\tilde{t}/2} \ r^{3-(\Delta+\tilde{\Delta})} \nonumber\\
&& \times \ \sqrt{s}\left(1+\frac{M^2R^4}{r^2(\tilde{\Delta}-3)^2}\right) (\sqrt{s-4M^2}+\sqrt{s})\left(\alpha'\tilde{s}\right)^{1+\alpha'\tilde{t}/2}e^{2-\alpha'\tilde{t}/2}\frac{\Gamma(-\alpha'\tilde{t}/4)}{\Gamma(1+\alpha'\tilde{t}/4)}   \, .
 \end{eqnarray}
Considering the limit $s>>M^2$ we obtain:
\begin{eqnarray}
&&\mathcal{A}_{\text{4-Regge limit}}^{\text{glueball-fermion}}(s,t) \approx \frac{3i\pi g_s^2\alpha'^4(\Delta-2)(\tilde{\Delta}-2)}{256\Lambda^{4-(\Delta+\tilde{\Delta})}R^{14-2(\Delta+\tilde{\Delta})}}I_{\tilde{\Delta}_1,\Delta_2,\Delta_3,\tilde{\Delta}_4}\int_{r_0}^{\infty} dr \ 4^{-\alpha'R^2t/2r^2}r^{3-(\Delta+\tilde{
   \Delta})}\nonumber\\
&& \times \ s \left(1+\frac{M^2R^4}{r^2(\tilde{\Delta}-3)^2}\right) \left(\alpha'R^2s/r^2\right)^{1+\alpha'R^2t/2r^2}e^{2-\alpha'R^2t/2r^2}\frac{4r^2}{\alpha'R^2|t|} \, .
\end{eqnarray}
The radial integral can be solved using the saddle-point approximation. Therefore, it is convenient to rewrite the integral as follows:
 \begin{eqnarray}
&&\mathcal{A}_{\text{4-Regge limit}}^{\text{glueball-fermion}}(s,t) \approx \frac{12 i \pi g_s^2 \alpha'^3 (\Delta-2)(\tilde{\Delta}-2)s}{256\Lambda^{4-(\Delta+\tilde{\Delta})}R^{16-2(\Delta+\tilde{\Delta})}|t|} \ I_{\tilde{\Delta}_1,\Delta_2,\Delta_3,\tilde{\Delta}_4} \times \nonumber\\
&&  \int_{r_0}^{\infty} dr \ \exp\bigg[\log\bigg(4^{-\alpha'R^2t/2r^2}r^{5-(\Delta+\tilde{
   \Delta})}\left(1+\frac{M^2R^4}{r^2(\tilde{\Delta}-3)^2}\right) \left(\alpha'R^2s/r^2\right)^{1+\alpha'R^2t/2r^2} \nonumber \\
&& \times \ e^{2-\alpha'R^2t/2r^2} \bigg)\bigg] \, .
\end{eqnarray}
Then, we can apply the saddle-point approximation to the function:
 \begin{eqnarray}
     h(r)=\log\bigg( 4^{-\alpha'R^2t/2r^2}r^{5-(\Delta+\tilde{
   \Delta})}(\alpha'R^2s/r^2)^{1+\alpha'R^2t/2r^2} \left(1+\frac{M^2R^4}{r^2(\tilde{\Delta}-3)^2}
   \right)e^{2-\alpha'R^2t/2r^2}\bigg) \, . \nonumber
 \end{eqnarray}
The following inequality is satisfied: 
 \begin{eqnarray}
     \frac{M^2R^4}{r^2(\tilde{\Delta}-3)^2}<\frac{M^2R^4}{r_0^2(\tilde{\Delta}-3)^2}=\frac{M^2R^4}{\Lambda^2R^4(\tilde{\Delta}-3)^2} \, .
 \end{eqnarray}
Recall that $\Lambda\sim M$ and $\tilde{\Delta} \equiv 2 \times \tilde{\Delta}^{(6)} =2 \times 7/2=7$, which corresponds to the sum of the scaling dimensions of two single-trace fermionic operators of the type ${\cal {O}}^{(6)}_k(x)$ with $k=0$, i.e. the minimum twist $\tilde{\tau}=3$, of the ${\cal {N}}=4$ SYM theory. Thus, we obtain:
 \begin{eqnarray}
    \frac{M^2R^4}{r^2(\tilde{\Delta}-3)^2}<\frac{1}{16} \, .
 \end{eqnarray}
Therefore, we find that the function $h(r)$ can be approximated by:
 \begin{eqnarray}
h(r)\approx \log\bigg( 4^{-\alpha'R^2t/2r^2}r^{5-(\Delta+\tilde{
   \Delta})}(\alpha'R^2s/r^2)^{1+\alpha'R^2t/2r^2} e^{2-\alpha'R^2t/2r^2}\bigg)     \, ,
 \end{eqnarray}
where the saddle-point condition $\frac{dh(r*)}{dr}=0$ implies that:
 \begin{eqnarray}
     r*\approx \sqrt{\frac{\alpha'|t|R^2\log(s/|t|)}{\Delta+\tilde{\Delta}-3}} \, .
 \end{eqnarray} 
Then, the dominant value of the radial coordinate is:
 \begin{eqnarray}
    r'=\text{max}\left(\Lambda R^2, \sqrt{\frac{\alpha'|t|R^2\log(s/|t|)}{\Delta+\tilde{\Delta}-3}}\right) \, ,
 \end{eqnarray}
which tells us that the Regge behavior manifests when 
 \begin{eqnarray}
      \sqrt{\frac{\hat{\alpha'}|t|\log(s/|t|)}{\Delta+\tilde{\Delta}-3}}<1 \, ,
 \end{eqnarray}
leading to
 \begin{eqnarray}
 \mathcal{A}_{\text{4-Regge limit}}^{\text{glueball-fermion}}(s,t)
  &\approx & \frac{12i\pi^{3/2} \alpha'^3 g_s^2(\Lambda R^2)^{5-(\Delta+\tilde{\Delta})} (\Delta-2)(\tilde{\Delta}-2)}{ 256\Lambda^{4-(\Delta+\tilde{\Delta})}R^{16-2(\Delta+\tilde{\Delta})}|t|}   \nonumber \\
&& \frac{2^{1/2-\frac{\alpha't}{\Lambda^2 R^2}}
  e^{2-\frac{\alpha't}{2 \Lambda^2 R^2}}s  \left(\frac{\alpha' s}{\Lambda^2 R^2}\right)^{1+\frac{\alpha' t}{2 \Lambda^2 R^2}}}{\sqrt{\left|\frac{3 \alpha' t \log \left(\frac{\alpha'
   s}{\Lambda^2 R^2}\right)+\alpha' t (2-3 \log4)+\Lambda^2 R^2 (\Delta+\tilde{\Delta}-3)}{\Lambda^4 R^6}\right|}} \ I_{\tilde{\Delta}_1,\Delta_2,\Delta_3,\tilde{\Delta}_4}   \, ,
 \end{eqnarray}
which, considering that $\hat{\alpha'}=\frac{1}{\sqrt{4\pi g_s N_c}\Lambda^2}=\frac{\alpha'}{\Lambda^2 R^2}$, can be written as:
 \begin{eqnarray}
 \mathcal{A}_{\text{4-Regge limit}}^{\text{glueball-fermion}}(s,t)&\approx&  \frac{12 i \pi^{3/2} \Lambda^3 \alpha'^{3/2} (4\pi g_s N_c)^{5/4}(\Delta-2)(\tilde{\Delta}-2)}{256 \cdot 16 \pi^2 N_c^2 \hat{\alpha'}|t|}    \nonumber \\ 
&& \frac{2^{\hat{\alpha'}|t|+\frac{1}{2}} e^{2+\frac{\hat{\alpha'}|t|}{2}}  \left(\hat{\alpha'} s\right)^{2+\frac{\hat{\alpha'} t}{2}}}{ \sqrt{|-3 \hat{\alpha'} |t| \log \left(\hat{\alpha'} s \right)-\hat{\alpha'} |t| (2-3 \log4)+\Delta+\tilde{\Delta}-3|}}   \ I_{\tilde{\Delta}_1,\Delta_2,\Delta_3,\tilde{\Delta}_4} \, . \nonumber \\
&&
 \end{eqnarray}
Phenomenologically we can set $\hat{\alpha'}|t|\sim 1$. Thus, we obtain the Regge behavior of the four-dimensional glueball-spin-1/2 fermion scattering amplitude:
 \begin{eqnarray}
     \mathcal{A}_{\text{4-Regge limit}}^{\text{glueball-fermion}}(s,t)&\approx & \frac{3i \Lambda^3 \alpha'^{3/2} (4\pi g_s N_c)^{5/4}(\Delta-2)(\tilde{\Delta}-2)   2^{3/2} e^{5/2}  \left(\hat{\alpha'} s\right)^{2+\frac{\hat{\alpha'} t}{2}}}{512\sqrt{\pi} N_c^2 \sqrt{|-3  \log \left(\hat{\alpha'}
   s\right)- (2-3 \log4)+\Delta+\tilde{\Delta}-3}|}  \nonumber \\
&& \times \  I_{\tilde{\Delta}_1,\Delta_2,\Delta_3,\tilde{\Delta}_4}   \, .
 \end{eqnarray}
The exponent $2+\hat{\alpha'} t/2$ of $s$ comes from the exchange of a single Reggeized graviton from the dual string theory perspective.

%
%
\section{Discussion and conclusions}
%
%

In this work we have obtained the four-dimensional glueball-fermion scattering amplitude in the high-energy limit, considering the fixed-angle and the Regge-limit behavior. This calculation is the first one focusing on the spin-1/2 fermion-glueball elastic scattering at high energy within the gauge/string theory duality framework. We carry out this from the closed-string scattering amplitude for two dilatons and two dilatinos in type IIB superstring theory developed in our work \cite{Martin:2025ije}. 

For the high-energy fixed-angle result given by equation (\ref{A4final}) the leading and subleading contributions are the following ones. Firstly, the leading contribution is proportional to:
\begin{eqnarray}
s^{5/2-\mathcal{T}/2}(1+\cos\theta)
\, , \label{dominant}
\end{eqnarray}
which dominates the scattering amplitude at any fixed angle with the exception of the backward scattering process. This contribution (\ref{dominant}) corresponds to the situation where the scattering amplitude behaves as $\mathcal{A} \propto s^{2-n/2} \ s^{L/2-1/2}$ with $n=\mathcal{T}$ and $L=2$ where, as commented in section 3, $L$ indicates the number of pairs of constituents which corresponds to distinct hadrons which have a large-angle scale-invariant interaction. Recall that this is a result obtained for the strongly coupled ${\cal {N}}=4$ SYM theory. On the other hand, from the point of view of perturbative non-Abelian gauge field theory calculations as in references \cite{Brodsky:1973kr,Brodsky:1974vy} $L>1$ is interpreted as the multiple scattering of several pairs of constituents, leading to non-planar contributions which are associated with the Landshoff's mechanism \cite{Landshoff:1974ew}. There is a subleading contribution to the glueball-fermion scattering amplitude of the form:
\begin{eqnarray}
\frac{1}{\hat{\alpha}'^{1/2} \ \lambda_{\text{'t Hooft}}^{1/4}} s^{2-\mathcal{T}/2}(\cos\theta-1) \, . \label{subdominant}
\end{eqnarray}
In this case $L=1$, which in the parton language means that this contribution to the glueball-fermion scattering amplitude only involves the scattering of a single constituent pair given by one constituent from the glueball and the other from the spin-1/2 fermion. However, there is a subtlety which comes from the fact that for the scattering angle close to $\pi$ the factor $(\cos\theta+1)$ of equation (\ref{dominant}) becomes close to zero. Thus, in the backward scattering limit the term proportional to $s^{2-\mathcal{T}/2}$ becomes the dominant one. Also, notice the suppression by the factor $\lambda_{\text{'t Hooft}}^{-1/4}$. This behavior corresponds to the strongly coupled gauge field theory.

We may consider the scattering of glueballs and fermions of spin-1/2 states created by single-trace ${\cal {N}}=4$ SYM operators with the lowest possible twists. The referred ${\cal {N}}=4$ SYM operators are ${\cal {O}}^{(8)}_0(x)$ of the form ${\text{Tr}}(F^2_+)$ with twist $\tau=\Delta_{\text{glueball}}=4$ and ${\cal {O}}^{(6)}_0(x)$ of the form ${\text{Tr}}(F_+ \lambda_{{\cal {N}}=4})$ with twist $\tau=\Delta_{\text{fermion}}-1/2=3$ \cite{DHoker:2002nbb}. In these operators the trace is taken on the adjoint representation of the gauge group $SU(N_c)$. They are gauge invariant operators built out of the ${\cal {N}}=4$ SYM gauge multiplet $(A_\mu^a,\lambda_{{\cal {N}}=4}, X_I)$, composed by the gauge fields $A_\mu^a$ (with the four-dimensional Lorentz indices $\mu=0, \dots, 3$ and the non-Abelian gauge group indices $a=1, \dots, N_c^2-1$), the Weyl fermions $\lambda_{{\cal {N}}=4}$, and the 6 real scalars $X_I$ (with $I=1, \dots, 6$). $F_+$ is the self-dual gauge-field strength. In this case we obtain the result given in equation (\ref{sigma-11}). An interesting point to be worked out yet concerns the result of the angular integral from (\ref{A4final}), which in this case is $I_{\tilde{\Delta}_1,\Delta_2,\Delta_3,\tilde{\Delta}_4} \equiv I_{7/2, 4, 4, 7/2}$, that we calculate next considering equation (\ref{angular-integral}): 
\begin{equation}
I_{7/2, 4, 4, 7/2} = \int_{S^5} d\Omega_5 \  \sqrt{\hat{g}_{S^5}}   \ Y^{(0)^*}_{(0,0,0,0,0)}(\Omega_5) \ Y^{(0)}_{(0,0,0,0,0)}(\Omega_5) \  Y^{(1/2)\dagger}_{(0,0,0,0,0)}(\Omega_5) \ Y^{(1/2)}_{(0,0,0,0,0)}(\Omega_5)  \,  , \label{example-1-angular-integral}
\end{equation}
where $\sqrt{\hat{g}_{S^5}}= \sin^4\theta_5 \sin^3\theta_4 \sin^2\theta_3 \sin\theta_2$ for the usual parametrization of the unit five-sphere. The scalar spherical harmonic is \cite{Martin:2024jpe}:
\begin{equation}
Y^{(0)}_{(0,0,0,0,0)}(\Omega_5) = \frac{1}{\pi^{3/2}} \, ,
\end{equation}
where the notation $(0,0,0,0,0) = (l_5, l_4, l_3, l_2, l_1)$ with $l_5 \geq l_4 \geq l_3 \geq l_2 \geq l_1$, and $l_5=k$, where $k$ counts the number of real scalar fields of the operator ${\text{Tr}}(F_+^2 X_I^k)$ \cite{Jorrin:2020cil,Jorrin:2020kzq}. In the case of the operator ${\cal {O}}^{(8)}_{k=0}$ we have $\Delta=\tau=4$.

Spinor spherical harmonics on the five-sphere were explicitly obtained in \cite{Jorrin:2020cil,Jorrin:2020kzq}
following the formalism developed in \cite{Camporesi:1995fb}. The spinor spherical harmonic with the minimal twist has a degeneration four, which is related to the lowest  irreducible representation $[1,0,0]$ of the $SU(4)_R$ $R$-symmetry group in the dual gauge field theory. Thus, there are four spin-1/2 fermionic modes $\lambda^-_{0_a}$ with $a=1, 2, 3, 4$. We refer the interested reader to references \cite{Jorrin:2020cil,Jorrin:2020kzq} for more details of this construction. We have checked that the final result does not depend on the choice of the initial state among the four fermionic modes belonging to the $[1,0,0]$ irreducible representation of $SU(4)_R$. Therefore, without loss of generality we may consider the lowest-twist ($\tau=3$) spherical harmonic: 
\begin{equation}
    Y^{(1/2)}_{(0,0,0,0,0)_{a=1}}= 
\frac{e^{-i{\cal{Q}} \theta_1}}{\pi^{3/2}} 
\begin{bmatrix} e^{-i \frac{1}{2}( \theta_3 -\theta_5)} 
\cos (\frac{\theta_2}{2}) \cos (\frac{\theta_4}{2}) \\ 
-e^{i\frac{1}{2}( \theta_3 +\theta_5)}\sin (\frac{\theta_2}{2}) 
\cos (\frac{\theta_4}{2}) \\ -e^{-i\frac{1}{2}( \theta_3 
+\theta_5)}\cos (\frac{\theta_2}{2}) \sin (\frac{\theta_4}{2}) 
\\e^{-i\frac{1}{2}( -\theta_3 +\theta_5)} \sin (\frac{\theta_2}{2}) 
\sin (\frac{\theta_4}{2}) \end{bmatrix} \, .
\end{equation}
This spinor spherical harmonic has the charge ${\cal{Q}}=\frac{1}{2}$ and also the sub-index $a$ related to the $[1,0,0]$ irreducible representation of $SU(4)_R$. We set $a=1$. Thus, it is very easy to show that $ Y^{(1/2)\dagger}_{(0,0,0,0,0)}(\Omega_5) \ Y^{(1/2)}_{(0,0,0,0,0)}(\Omega_5) = 1/\pi^3$ and therefore the integral (\ref{example-1-angular-integral}) gives:
\begin{equation}
I_{7/2, 4, 4, 7/2} = \frac{1}{\pi^3} \,  .
\end{equation}

We may consider the case of ${\cal {O}}^{(8)}_1(x)$ of the form ${\text{Tr}}(F^2_+ X_I)$ with twist $\tau=\Delta_{\text{glueball}}=5$ and again take ${\cal {O}}^{(6)}_0(x)$ of the form ${\text{Tr}}(F_+ \lambda_{{\cal {N}}=4})$ with twist $\tau=\Delta_{\text{fermion}}-1/2=3$. Then, the angular integral to solve is $I_{\tilde{\Delta}_1,\Delta_2,\Delta_3,\tilde{\Delta}_4} \equiv I_{7/2, 5, 5, 7/2}$. In this case we have the following 6 orthonormalized scalar spherical harmonics:
\begin{eqnarray}
Y_{(1,0,0,0,0)}^{(0)} &=&  \frac{\sqrt{6}}{\pi^{3/2}} \ \cos\theta_5 \ , \\
Y_{(1,1,0,0,0)}^{(0)} &=&  \frac{\sqrt{6}}{\pi^{3/2}} \ \cos\theta_4 \ \sin\theta_5 \ , \\
Y_{(1,1,1,0,0)}^{(0)} &=&  \frac{2\sqrt{6}}{\pi^{3/2}} \ \frac{ \cos\theta_3 \ \sin\theta_4 \ \sin^2\left(\frac{\theta_5}{2}\right) \ (1+\cos\theta_5)}{\sin\theta_5} \, , \\
Y_{(1,1,1,1,0)}^{(0)} &=&  \frac{2\sqrt{6}}{\pi^{3/2}} \ \frac{ \cos\theta_2 \ \sin\theta_3 \ \sin\theta_4 \ \sin^2\left(\frac{\theta_5}{2}\right) \ (1+\cos\theta_5)}{\sin\theta_5} \, , \\
Y_{(1,1,1,1,1)}^{(0)} &=&  \frac{2\sqrt{3}}{\pi^{3/2}} \ e^{-i \theta_1} \ \frac{ \sin\theta_2 \ \sin\theta_3 \ \sin\theta_4 \ \sin^2\left(\frac{\theta_5}{2}\right) \ (1+\cos\theta_5)}{\sin\theta_5} \, , \\
Y_{(1,1,1,1,-1)}^{(0)} &=&  \frac{2\sqrt{3}}{\pi^{3/2}} \ e^{i \theta_1} \ \frac{ \sin\theta_2 \ \sin\theta_3 \ \sin\theta_4 \ \sin^2\left(\frac{\theta_5}{2}\right) \ (1+\cos\theta_5)}{\sin\theta_5} \, .
\end{eqnarray}
The results of the angular integrals in all these cases give also $1/\pi^3$. On the other hand, the leading contribution to the differential cross section at fixed angle is:
\begin{eqnarray}
\frac{d\sigma}{dt} \propto s^{-13}f(|t|/s) \, ,
\end{eqnarray}
which also has $L=2$. From the orthogonality of the spherical harmonics it is possible to derive selection rules leading to elastic scattering of the ${\cal {N}}=4$ SYM states created by the fermionic operator ${\cal {O}}^{(6)}_0(x)$ with any scalar operator ${\cal {O}}^{(8)}_k(x)$ with $k \ge 0$, preserving the value of $k$ of the glueballs.

At this point it is interesting to make some comments on Landshoff's mechanism \cite{Landshoff:1974ew} since it is related to multiple scattering of partons in processes such as hadron hard scattering and deep inelastic scattering among other high-energy scattering processes. For the elastic amplitude at fixed-angle, the scattering amplitude is proportional to $s^{2-{\cal {T}}/2} s^{L/2-1/2}$, considering $L$ pairs of partons from different hadrons as explained before, which in principle can become the leading contribution. However, perturbative higher-order corrections imply that the behavior could change not so drastically, and the result would be close to the case of a single parton-pair scattering, i.e. $L=1$, somehow obliterating the multiple parton-pair scattering contributions \cite{Mueller:1981sg}.
The very large number of Feynman diagrams emerging from QCD makes the calculations beyond the simple quark counting rules very difficult \cite{Dremin:2012ke}.

In addition, in section 4 we have obtained the Regge limit of the glueball-fermion scattering amplitude leading to the exchange of a single Reggeized graviton in the dual string theory description. For comparison, let us recall that in the case of four glueballs the Regge limit also leads to the propagation of a single Reggeized graviton. In the case four spin-1/2 fermions we have obtained the reggeization of a graviton, and also a sub-leading contribution given by the reggeization of a vector field which turns out to be a linear combination of off-diagonal fluctuations of the metric tensor and vector fluctuations of the Ramond-Ramond four-form field potential.

~

~

%
\centerline{\large{\bf Acknowledgments}}
%

\vspace{0.5cm}

The work of L.M., M.P. and M.S. has been supported by the Consejo Nacional de Investigaciones Cient\'{\i}ficas y T\'ecnicas of Argentina (CONICET).

\pagebreak

\end{document}